\documentclass[twocolumn,showpacs,preprintnumbers,amssymb,pra]{revtex4}

\usepackage{amssymb}
\usepackage{amsmath}
\usepackage{graphicx}
\usepackage{dcolumn}
\usepackage{bm}
\usepackage{verbatim}
\usepackage{array}
\usepackage{epsfig}
\usepackage{amsbsy}
\usepackage{color}
\usepackage{soul}

\def\sprm#1#2{  \left\langle #1 \left\vert \right. #2 \right\rangle   }
\def\rmem#1#2#3{  \left\langle #1 \left\vert \left\vert  #2
                  \right\vert \right\vert #3 \right\rangle   }

\newcommand{\ds}{\displaystyle}
\newcommand{\ddsum}[1]{{\displaystyle \sum_{ #1 }}}

\newcommand{\supercomas}[1]{``#1''}

\def\bra#1{\mathinner{\langle{#1}|}}
\def\ket#1{\mathinner{|{#1}\rangle}}
\newcommand{\braket}[2]{\langle #1|#2\rangle}

\newcommand{\dd}{\mathrm{d}}
\newcommand{\ti}{~}

\begin{document}
\preprint{\hfill\parbox[b]{0.3\hsize}{ }}

\title{Parametrization of the angular correlation and degree of linear polarization in two-photon decays
 of hydrogen-like ions}

\author{P.\ Amaro$^{1,2 } \footnote{pamaro@physi.uni-heidelberg.de}$,
        F.\ Fratini$^{1,3,4}$,
        S.\ Fritzsche$^{3,4}$,
        P.\ Indelicato$^{5}$,
        J.\ P.\ Santos$^{2}$,
        A.\ Surzhykov $^{1,3}$
          }
\affiliation
{\it
$^{1}$ Physikalisches Institut, Universit\"at Heidelberg, D-69120 Heidelberg, Germany \\
$^{2}$
Centro de F\'isica At\'omica, CFA, Departamento de F\'isica, Faculdade de Ci\^encias e Tecnologia,
 FCT, Universidade Nova de Lisboa, 2829-516 Caparica, Portugal \\  
$^3$ 
GSI Helmholtzzentrum f\"ur Schwerionenforschung, D-64291 Darmstadt,
 Germany \\
$^4$  
Department of Physics, P.O. Box 3000, Fin--90014 University
 of Oulu, Finland\\
$^5$
Laboratoire Kastler Brossel, \'Ecole Normale Sup\'erieure, CNRS,  
Universit\'e P. et M. Curie -- Paris 6, Case 74; 4, place Jussieu, 75252 Paris CEDEX 05, France}

\date{Received: \today  }


\begin{abstract}

The two-photon decay in hydrogen-like ions is investigated within the framework of second order perturbation theory and Dirac's relativistic equation. Special attention is paid to the angular correlation of the emitted photons as well as to the degree of linear polarization of one of the two photons, if the second is just observed under given angles. Expressions for the angular correlation and the degree of linear polarization are expanded in terms of $\cos\theta$-polynomials, whose coefficients depend on the atomic number and the energy sharing of the emitted photons. The effects of including higher (electric and magnetic) multipoles upon the emitted photon pairs beyond the electric-dipole approximation are also discussed. Calculations of the coefficients are performed for the transitions $2s_{1/2}\rightarrow1s_{1/2}$, $3d_{3/2}\rightarrow1s_{1/2}$ and $3d_{5/2}\rightarrow1s_{1/2}$, along the entire hydrogen isoelectronic sequence ($1\le Z \le 100$).
\end{abstract}


\maketitle


\section{Introduction}
\label{intro}

Studies on the two-photon decay of atoms and ions have a long tradition, which can be traced back to G\"oppert-Mayer in 1931 \cite{GoM31}. 
These studies made the first prediction of non-linear (second-order) processes in atomic physics and were originally focused upon the total transition rates and the spectral distribution of the two-photon emission \cite{BrT40, DuB93, Au76,AlA97, ScM99, MoD04, MoD042, IlU06,  GoD81, ToL90, DeJ97, SaP98,KuT09,TrS09}. %
More recently, various investigations on such processes have aimed to test not only the standard model by measuring parity
nonconservation (PNC) effects, but also the basics of quantum theory, using two-photon decay channels (see \cite{pdb1985, ssi1989, KlD97,  FrI05, QuantumInf, PNC, SIS11} and references therein). 
Recent technical advances in polarization and position-sensitive detectors have opened up the possibility of investigating angular and polarization properties of the radiation emitted in atomic decays~\cite{RSI67,TaS06, StS07}. 
For medium-- and high--$Z$ ions, in particular, measurements of two-photon angular and polarization properties are presently planned to be performed at GSI in Darmstadt in forthcoming years. \\
To support these measurements and to extend previous investigations \cite{SuK05, SuV10, Hyperfine}, we here present a relativistic study on the \emph{angle correlation} and \emph{degree of linear polarization}  of the radiation emitted in two-photon decays of hydrogen--like ions.  
We investigate both, the angular dependence of the differential decay rate (referred to as angular correlation) as well as the degree of linear polarization of the \supercomas{first} photon, if  the \supercomas{second} photon is observed under certain angles and if its polarization properties remain unobserved.
Moreover, by using the symmetry properties of the second order transition amplitude, the angular correlation and the degree of linear polarization of the emitted photons are written in terms of polynomial expansions in $\cos\theta$,  where $\theta$ is the (opening) angle between the direction of the two photons. Such a (relativistic) \emph{parametrization} can be readily compared with other theoretical calculations and provides a theoretically well-justified fit model for future experiments that involve two-photon transitions.
Up to now, there is no \emph{fully}-relativistic parametrization of such quantities for hydrogen-like ions available. A previous parametrization of the angular correlation performed by Au~\cite{Au76}  was restricted to the $2s_{1/2}\rightarrow1s_{1/2}$ transition as well as to a non-relativistic framework.  

This paper is organized as follows: in Sec.~\ref{two_pho_Am} we give a brief overview of the background
theory involved in two-photon emission and introduce the (second-order) reduced amplitudes, which represent the building blocks of the theoretical calculation. Next, we define the angle correlation and degree of linear polarization functions together with its parametrization procedure in Secs.~\ref{Angle_angle} and \ref{Pola_angle}. The numerical evaluation of the reduced amplitudes is described in Sec.~\ref{S_elemete}.
Section~\ref{Resul} contains the coefficients of the expansions for the $2s_{1/2}\rightarrow1s_{1/2}$, $3d_{3/2}\rightarrow1s_{1/2}$ and $3d_{5/2}\rightarrow1s_{1/2}$  transitions. 
In order to describe more easily the relativistic behavior of these coefficients, we express them in terms of reduced two-photon matrix elements. In addition, the effect of higher order multipoles other than the dominant electric dipole in these evaluations is also discussed.
A brief summary is given in Sec.~\ref{sum}.


\section{Theory}
\label{theory}


\subsection{\label{two_pho_Am} Two-photon transition amplitude}
The standard theoretical formalism for the description of two--photon decays in atoms or ions is based on the relativistic second-order perturbation theory and has recently been applied in Refs.~\cite{SuK05, SuV10, Hyperfine,GoD81}.  In this paper, we therefore restrict ourselves to a compilation of those equations that  are needed for the present analysis.
The second--order transition amplitude for the emission of two photons with wave vectors $\bm{k}_i$ ($i$ = 1, 2) and polarization vectors $\bm{u}_{\lambda_i}$ ($\lambda_i = \pm 1$ ), from an initial atomic state $\ket{i}=\ket{n_i j_i m_i}$ to a final atomic state $\ket{f}=\ket{n_f j_f m_f}$, with well-defined principal quantum number $n_{i,f}$, total angular momenta $j_{i,f}$ and their projections $m_{i,f}$, is given by \cite{GoD81}
\begin{equation}
\begin{array}{l}
\ds\mathcal{M}_{fi}^{\bm k_1\bm k_2}(m_i, m_f ,\lambda_1,\lambda_2 )= \\
\quad\ds \ddsum{\nu}\!\!\!\!\!\!\!\!\int
\Big[
\frac{ \bra{f} \hat{\mathcal{R}}^{\dag}(\bm k_1, \lambda_1)
\ket{\nu}\bra{\nu} \hat{\mathcal{R}}^{\dag}(\bm k_2, \lambda_2) \ket{i}}{E_{\nu}
-E_i+\omega_2 } \\[0.4cm]
\qquad\qquad \ds + \left( 1 \longleftrightarrow 2\right) \Big] ~ .
\end{array}
\label{Mfi}
\end{equation}
Here, $\omega_{1,2}$ are the energies of the emitted photons and  the transition operator $\hat{\mathcal{R}}(\bm k, \lambda)$ in Eq.~(\ref{Mfi}) denotes the interaction between the electron and the electromagnetic radiation. 
Moreover, the intermediate states $\ket{\nu}=\ket{n_\nu j_\nu m_\nu}$ in Eq.\eqref{Mfi} form a complete set of states, including both discrete and positive/negative energy eigenvalues of the Dirac spectrum. The symbol $\ds\ddsum{\nu}\!\!\!\!\!\!\!\!\int$ stands for a summation over the discrete as well as an integration over the continuum part of this intermediate states.
In the soÐ-called Coulomb gauge, which corresponds to the velocity form of the electron-photon interaction operator in the non-relativistic limit, 
the transition operator reads
\begin{eqnarray}
\label{R_interaction}
\hat{\mathcal{R}}(\bm k, \lambda)  =  \bm\alpha \cdot \bm u_{\lambda} e^{i \bm k \cdot\bm r} \, ,
\end{eqnarray}
where $\bm\alpha$ is the usual vector of Dirac matrices.

Owing to the conservation of energy,  the initial and final ionic energies $E_i$ and $E_f$ are related to the energies of the emitted photons $\omega_{1,2}$ by
\begin{equation}
E_i - E_f = \omega_{1} + \omega_{2} \, .
\label{encons}
\end{equation}
Instead of working with photon energies, it is therefore more convenient to describe the decay using the energy sharing parameter $y =\omega_1/(\omega_1 + \omega_2)$.

The evaluation of the angular and polarization properties of the emitted radiation requires a spherical tensor decomposition of the photon fields contained in Eq.~(\ref{Mfi}) into their electric and magnetic multipole components \cite{Rose}. This decomposition reads
\begin{eqnarray}
\label{A_decomposition}
\bm u_{\lambda} e^{i \bm k \cdot\bm r} 
&=& \sqrt{2 \pi} \sum\limits_{L_\gamma = 1}^{\infty} \sum\limits_{M_\gamma = -L_\gamma}^{L_\gamma} \sum\limits_{p = 0, 1}
i^{L_\gamma} \, [L_\gamma]^{1/2} \, (i \lambda)^p \, \nonumber \\
&\times& \hat{a}^{p}_{L_\gamma M_\gamma}(k, \bm{r}) \, D^{L_\gamma}_{M_\gamma \lambda}(\bm{\hat k}) \, ,
\end{eqnarray}
where $[L_\gamma] = 2L_\gamma + 1$, $k=|\bm k|$, $D^{L_\gamma}_{M_\gamma \lambda}$ are the Wigner rotation matrices of rank  $L_\gamma$ and $\hat{a}^{p=0,1}_{L_\gamma M_\gamma}(k,\bm{r})$ refer to the magnetic ($p$ = 0) and the electric ($p$ = 1) components.  Note that the index $\gamma$ stands to the first ($\gamma=1$) or second ($\gamma=2$) photon.
The explicit form of the quantities $\hat{a}^{p}_{L_\gamma M_\gamma}(k,\bm{r})$ can be found in Ref.~\cite{GoD81}. 
Thus, for example, the term with $L_\gamma=1$ and $p=1$ is referred as an electric-dipole ($E1$), while the other designate higher (magnetic and electric) multipoles.\\
Inserting Eqs.~(\ref{R_interaction}--\ref{A_decomposition}) into Eq.~(\ref{Mfi}) and using the Wigner--Eckart theorem, we get the general expression for the second-order transition amplitude
\begin{widetext}
\begin{eqnarray}
\label{M_amplitude_new}
\ds\mathcal{M}_{fi}^{\bm k_1\bm k_2}(m_i, m_f ,\lambda_1,\lambda_2)
&=& 2 \pi \,
\sum\limits_{L_1 M_1 p_1} \sum\limits_{L_2 M_2 p_2}
(-i)^{L_1 + L_2} \, [L_1, L_2]^{1/2} \, (-i \lambda_1)^{p_1} \,
(-i \lambda_2)^{p_2} \, D^{L_1 *}_{M_1 \lambda_1}(\bm{\hat{k}}_1) \,
D^{L_2 *}_{M_2 \lambda_2}(\bm{\hat{k}}_2) \nonumber \\
&\times& \sum\limits_{j_\nu m_\nu}
\frac{1}{[j_i, j_\nu]^{1/2}}
\Bigg[
\sprm{j_f m_f \, L_1 M_1}{j_\nu m_\nu} \, \sprm{j_\nu m_\nu \, L_2 M_2}{j_i
m_i} \,
S_{L_1 p_1, \, L_2 p_2}^{j_\nu}(\omega_2) \nonumber \\
&+&
\sprm{j_f m_f \, L_2 M_2}{j_\nu m_\nu} \, \sprm{j_\nu m_\nu \, L_1 M_1}{j_i m_i} \,
S_{L_2 p_2, \, L_1 p_1}^{j_\nu}(\omega_1) \Bigg] \,
,
\end{eqnarray}
where the reduced amplitudes $S^{j_\nu}_{L_1 p_1, \, L_2 p_2}$ are defined as
\begin{eqnarray}
\label{S_one_electron}
S^{j_\nu}_{L_1 p_1, \, L_2 p_2}(\omega_2) =
\sum\limits_{n_\nu}
\frac{\rmem{n_f j_f}{\bm{\alpha}  \,
\hat{a}^{p_1 \dag}_{L_1}(k_1)}{n_\nu j_\nu}
\rmem{n_\nu j_\nu}{\bm{\alpha}  \, \hat{a}^{p_2 \dag}_{L_2}(k_2)}{n_i j_i}
}{E_\nu - E_i + \omega_2} \, .
\end{eqnarray}
\end{widetext}
The numerical evaluation of expression (\ref{S_one_electron}) is rather difficult and has been accomplished using several different methods \cite{GoD81, PRL29, JPB31, JPA24, SaJ96, ShT04}.
The main difficulty of this evaluation lies in the summation over the infinite intermediate one particle Dirac states $\ket{n_{\nu},j_{\nu}}$, which runs over the discrete as well as the positive and negative continuum part of the spectrum.  
In this work, Eq.~(\ref{S_one_electron}) is evaluated by exploiting both the relativistic Coulomb-Green's function \cite{JPB31, JPA24} and the discrete basis set approaches \cite{SaP98, SaJ96, ShT04}. We outline the theory of these two approaches in Sec.~\ref{S_elemete}.
In the following sections, we define the angular correlation function and the degree of linear polarization of the radiation emitted in two-photon decays of hydrogen-like ions, which are the two observables of interest in the present work.


\subsection{Angular correlation function}
\label{Angle_angle}

\begin{figure}
\includegraphics[width=\columnwidth ]{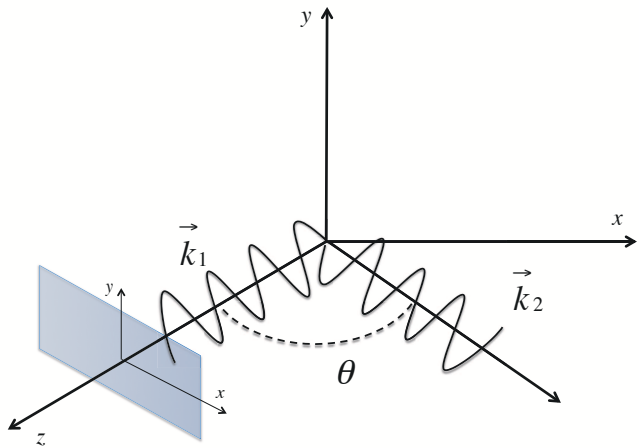}
\caption{(Color online) Geometry of the two-photon decay. The $z$ axis is collinear with the first photon momentum vector $\bm{k_1}$. The plane $zx$ contains the second photon momentum vector $\bm{k_2}$ and is refereed as the reaction plane. The angle $\theta$ is the opening angle between the direction of the two photons.}
\label{fig1}
\end{figure}

If we assume that the excited ions are initially unpolarized and that the spin states of the emitted photons remain unobserved during the measurement, 
the differential decay rate can be written in atomic units as \cite{GoD81}
\begin{equation}
\begin{array}{l c l}
\ds \frac{\dd \bar{W}}{\dd \omega_1\,\dd \Omega_1\, \dd \Omega_2} &=&\ds \frac{\omega_1\omega_2}{(2\pi)^3c^2}\frac{1}{2j_i+1} \\ [0.4cm]
&& \ds \times \ddsum{m_i,m_f}\ddsum{\lambda_1,\lambda_2}\Big|
\ds\mathcal{M}_{fi}^{\bm k_1\bm k_2}
(m_f,m_i,\lambda_1,\lambda_2)\Big|^2 .
\label{rate}
\end{array}
\end{equation}
Here, $\mathcal{M}_{fi}$ is given by Eq.~(\ref{Mfi}). 
To further evaluate the amplitude $\mathcal{M}_{fi}$, we must specify the geometry under which the photon emission is considered. Since there is no preferred direction for the decay of an unpolarized  ion, it is generally more convenient to adopt the quantization ($z$) axis along the momentum of the \supercomas{first} photon: $\vec k_1||z$ (see Fig.~\ref{fig1}). This choice of quantization axis enables us to define the angular correlation function as
%
%
\begin{eqnarray} 
\ds W^y(\theta) &=& 8\pi^2 (E_i-E_f) \nonumber \\
&&\times\ds \frac{\dd \bar{W}}{\dd \omega_1\,\dd \Omega_1\, \dd \Omega_2} 
(\theta_1=0,\phi_1=0,\phi_2=0) \nonumber \\
&=& \frac{\dd \bar{W}}{\dd y\,\dd \cos \theta }\, ,
\label{angcorr}
\end{eqnarray}
where we have defined the opening angle ($\theta$) as the polar angle of the \supercomas{second} photon ($\theta_2$) in order to simplify notation. In Eq.~\eqref{angcorr}, the factor $8\pi^2$ hereby arises from the integration over the solid angle of the first photon ($\dd\Omega_1$) as well as the integral taken over the azimuthal angle of the second photon ($\dd\phi_2$). Therefore, $W^y(\theta)$ represents the (density) number of photon pairs emitted per second for a given opening angle $\theta$ and energy fraction $y$, irrespectively of their polarizations. 

Due to the photon-photon permutation symmetry that characterizes the amplitude in Eq.~(\ref{Mfi}) and corresponds to the exchange of the emitted photons, the analytic expression for $W^y(\theta)$ can only contain terms that are $\theta$--even under the  algebraic replacement of $\theta\to -\theta$. We can thus decompose $W^y(\theta)$ in terms of a polynomial expansion of powers of $\cos\theta$
%
\begin{equation}
\begin{array}{l c l}
\ds W^y(\theta) &=& a_{0} \left(1 + \sum\limits_{i=1}^{\infty} a_i \cos ^i \theta \right)~,
\end{array}
\label{Acorr_expression}
\end{equation}
where the parameters $a_{i}$ depend on the atomic number of the ion, the energy sharing parameter $y$ and the states involved in the decay.  Note that only the parameter $a_0$ has units, which in S. I.  are s$^{-1}$.
We define the reaction plane as the plane spanned by the propagation directions of the photons. 
%

\subsection{Degree of linear polarization}
\label{Pola_angle}


We used the spin-polarization density matrix to analyze the polarization properties of the emitted radiation, which has been applied several times to the two-photon decay rate \cite{SuK05, Balashov}. 
The first Stokes parameter is given by \cite{Balashov}
\begin{equation}
\begin{array}{l c l}
P_{1 }(\theta ) &=&  \dfrac{ 2 }{\mathcal{N}} \textrm{Re} \left[  \ddsum{m_i,m_f, \lambda_2 } \ds\mathcal{M}_{fi}^{\bm k_1\bm k_2} (m_f,m_i,\lambda_1=1,\lambda_2) \right. \\ 
& & \hspace*{1.8 cm}  \left.   \ds \times \mathcal{M}_{fi}^{\bm k_1\bm k_2 \ast } (m_f,m_i,\lambda_1=-1,\lambda_2 ) \right] ,
\end{array}
\label{PL_def}
\end{equation}
where $\mathcal{N}$ is a normalization coefficient given by
\begin{equation}
\mathcal{N}=\ddsum{m_i,m_f}\ddsum{\lambda_1,\lambda_2}\Big|
\ds\mathcal{M}_{fi}^{\bm k_1\bm k_2}
(m_f,m_i,\lambda_1,\lambda_2)\Big|^2 ,
\end{equation}
which assures that the trace of the spin-polarization density matrix is unitary.
The derivation of Eq. (\ref{PL_def}) follows from the assumption that the polarization of only one (first) of the photons is observed. 
The first Stokes parameter of the first photon is also given by \cite{Balashov}
\begin{equation}
\begin{array}{l c l}
\ds P_{1}\left( \theta \right) &=& 
\ds\frac{
I_{\parallel}-I_{ \perp}
}{
I_{ \parallel}+I_{ \perp}}\; ,
\end{array}
\label{Pl}
\end{equation}
where $I_{ \parallel(\perp)}$  denotes the two-photon decay rate, if the second photon polarization remains unobserved and the linear polarization of the first photon is detected parallel (perpendicular) to the reaction plane. Note that the term $\mathcal{N}$ is proportional to $\ds W^y(\theta)$ defined in the previous section. Therefore, by employing Eq.~(\ref{Acorr_expression}) into Eq.~(\ref{PL_def}) and using the symmetry properties of $\mathcal{M}_{fi}$, we can parametrize $P_1$ as
\begin{equation}
\begin{array}{l}
\ds P_{1}\left(\theta\right)=
\frac{b_{0} \left(1 + \sum\limits_{k=1}^{\infty} b_k \cos ^k \theta \right)}{a_{0} \left(1 + \sum\limits_{i=1}^{\infty} a_i \cos ^i \theta \right) } \; .
\end{array}
\label{Pl_par}
\end{equation}
Like the case of $a_i$, parameters $b_i$ depend on the ion atomic number and energy sharing. 
It should be noted that within the current geometry and since we assume that the atom/ion is unpolarized, the second Stokes parameter vanishes. Therefore, the degree of linear polarization, which we define as $P_L(\theta)$, is equal to the absolute value of the first Stokes parameter, i.e., $P_L=\sqrt{P_1^2+P_2^2}=|P_1|$ \cite{Balashov}.

We find by analytical evaluation that the relations $b_{1}=-b_{3}$ and $b_2=-(1+b_4)$ 
are valid for the entire isoelectronic sequence, energy shares or type of transition. 
By considering these relations,  the degree of polarization $P_L(\theta)$ takes the form
\begin{equation}
\begin{array}{l}
\ds P_{L}\left(\theta\right)=
\left| \frac{b_{0} \big[ \sin^2 \theta \big(1 + b_{1}\cos\theta - b_{4}\cos^{2}\theta )+...\big]}{W^y(\theta) } \right| \; .
\end{array}
\label{Pl_par_co}
\end{equation}
It can observed that by neglecting parameters $b_{i}$ for $i>4$, the degree of polarization is zero for values of $\theta$ equal to $0^\circ$ or $180^\circ$. This is expected since for these angles there is no unique plane of reaction defined by the direction of the photons, neither a unique coordinated system that defines the first photon polarization. Further relations between $b_{i}$ with $i>4$ can thus be expected in order to have $P_L=0$ for these geometrical settings. Obtaining such relations goes beyond the scope of this work since as discussed in Sec.~\ref{Resul}, the contribution of these (higher) parameters to the degree of polarization can be neglected.
\begin{figure*}
\centering
\includegraphics[height=10cm, width=\textwidth ]{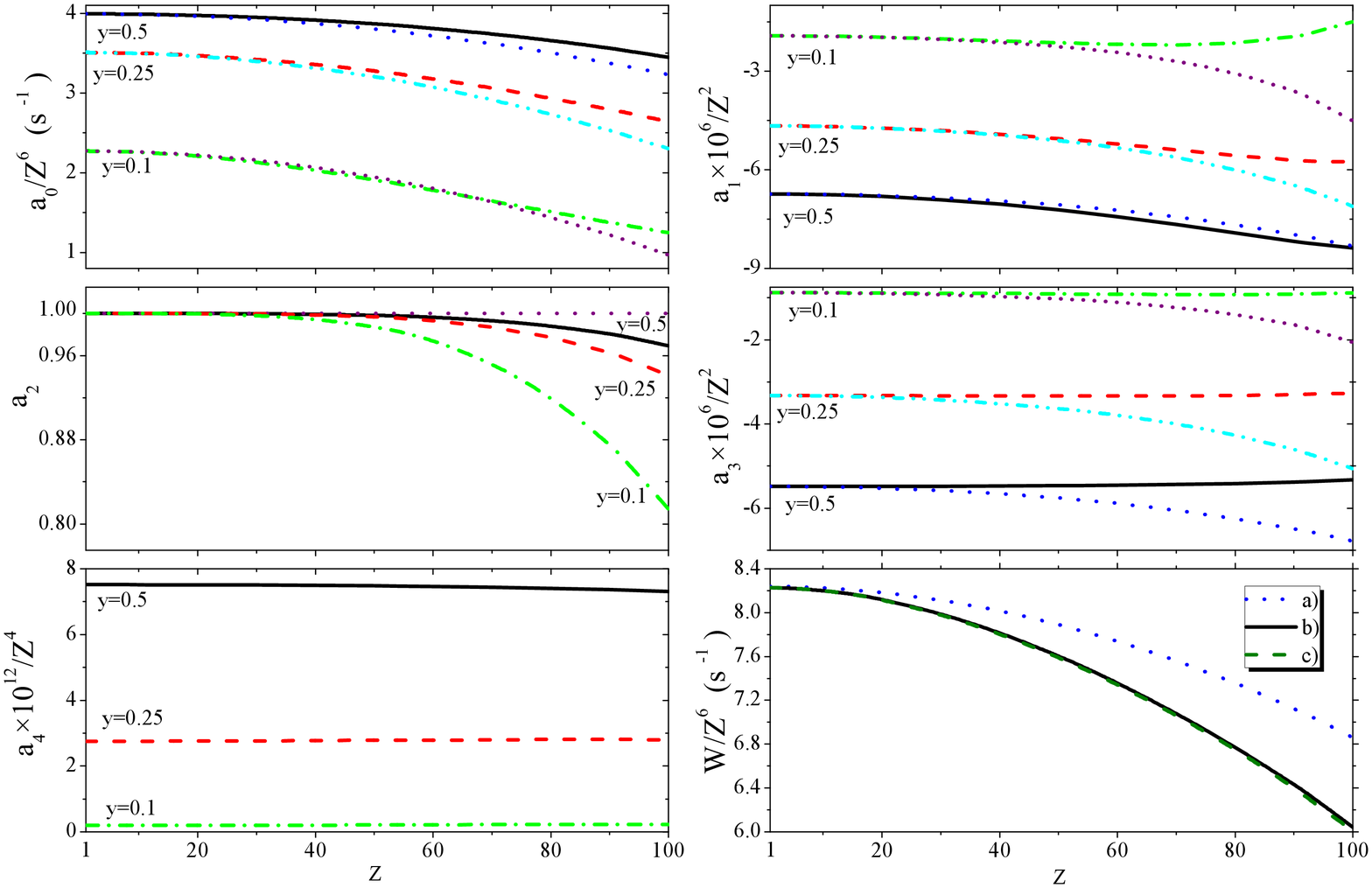}
\caption{(Color online) Values of the parameters $a_{i}$ defined in Eq.~(\ref{Acorr_expression}), for the $2s_{1/2}\rightarrow1s_{1/2}$ transition as a function of $Z$. The values obtained in this work, which correspond to different energy shares $y$ = 0.1, 0.25, 0.5 are represented by a black-solid, red-dashed and green-dot-dashed curve, respectively. With exception of $a_0$ (given in s$^{-1}$), all parameters are dimensionless. The same energy shares values given by Au~\cite{Au76} are represented by a blue dotted line ($y=0.5$), a cyan dot-dot-dashed line ($y=0.25$) and by the purple short dashed line ($y=0.1$). 
The parameters  $a_2$ of Ref.~\cite{Au76} are always equal to one, as represented by the purple dotted line. On the bottom right plot, the blue dotted line refers to values of the total decay rate obtained from the parametrization of Ref.~\cite{Au76}. 
The black line represents the total decay values obtained from our current parametrization. The dark green dashed  line contains the values provided by Refs.~\cite{GoD81,SaP98}. The level of precision of the plot does not allow to distinguish differences between these references. Values of parameters $b_{i}$ can be obtained from this figure using the relations $b_0\approx-a_0$, $b_1\approx a_3$ and $b_4\approx-a_4$. }
\label{fig2}
\end{figure*}


\subsection{Computation of the $S$ functions}
\label{S_elemete}
The reduced matrix elements in Eq.~(\ref{S_one_electron}) were evaluated by making use of both the relativistic \emph{Coulomb-Green's} function \cite{JPB31, JPA24} and the \emph{finite basis set} \cite{SaP98, SaJ96, ShT04} approaches. The Coulomb-Green's function is defined to satisfy the equation
\begin{equation}
(\hat {\bf H} - E)G_E(\bf{r},\bf{r}')=\delta(\bf{r}-\bf{r}') \;,
\label{eqGF}
\end{equation}
where $\hat {\bf H}$ is the Hamiltonian that describes the atom or ion. The Coulomb-Green's function can formally be written as
\begin{equation}
\ds G_E(\bf{r},\bf{r}')=\ddsum{\nu}\!\!\!\!\!\!\!\!\int \frac{\braket{\bf{r}}{\Psi_{\nu}}\braket{\Psi_{\nu}}{\bf{r}'}}{E_{\nu}-E} \;,
\label{SolFormally}
\end{equation}
By comparing Eq.~(\ref{Mfi}) with Eq.~(\ref{SolFormally}), we recognize that $G_E$ is an essential part of the transition amplitude for which care has to be taken with its numerical evaluation.
We note that $G_E$ is known analytically both for the non-relativistic Schr\"odinger-Coulomb as well as the relativistic Dirac-Coulomb Hamiltonian and it can be represented in terms of Laguerre polynomials \cite{JPA24}.

The second approach (finite basis set) has been recently applied to two-photon decay \cite{asp2009}, two-photon absorption  \cite{sis2011} and Rayleigh scattering \cite{saf2012}, in hydrogen-like ions. It is based on the supposition of enclosing the atom/ion in a potential box with finite (although large enough to get a good approximation) radius.
 These boundary conditions imply a \supercomas{discretization of the continuum}, which enables one to replace the infinite sum over the bound stats as well as the integration over the continuum by just a \supercomas{finite summation} over a basis set.
 
  In this work we use the \emph{B-splines} basis set. Another basis set, \emph{B-polynomials} \cite{asp2011}, was used to check the results of the finite basis set method. By carrying out the calculations, we verified that both approaches yielded identical results for the values of the coefficients in Eqs.\ti(\ref{Acorr_expression}) and (\ref{Pl_par}). 

Finally, to further assess the accuracy of the reduced matrix elements $S^{j_\nu}_{L_1 p_1, \, L_2 p_2}$, these elements were evaluated in both Coulomb and Lorentz gauges and tested in a total decay rate expression as performed in Ref.~\cite{SaP98}.


\section{Results and discussion}
\label{Resul}

Fig.~\ref{fig2} displays the values of the coefficients $a_i$ in the expansion (\ref{Acorr_expression}) of the two-photon correlation function $W^y(\theta)$ for the two-photon transition $2s_{1/2}\to1s_{1/2}$ as  function of the nuclear charge $Z$. 
We considered only multipoles ($L<4$) that gives a visible contribution to the figure. Several different energy shares ($y=0.1,~ 0.25, ~0.5$) are displayed.
For small atomic numbers, the coefficients $a_{i }$ with $i>2$ and $i=1$ nearly vanish, while $a_{2}$ tends to 1.
We notice that the angular correlation function is well described, in the low-Z regime, by $W^y(\theta)\sim a_0(1 + \cos^2\theta)$, which corresponds  to the dipole approximation \cite{Klarsfeld}. This is in agreement also with the more general result of Yang \cite{yang}, who predicts an angular correlation of the form $(1+\beta \cos^2(\theta))$ for dipole-dipole transitions. 
A deviation to this expression arise from $a_1$ and $a_3$, which leads to an asymmetry (with respect to $\theta=90^\circ$) and to a \supercomas{tilt} of the angular correlation function, i.e., a slight preference of a back-to-back emission of the two photons ($\theta=180^\circ$), when compared with an emission into the same direction ($\theta=0^\circ$). 
In order to quantify this deviation, we write the coefficients $a_i$ directly in terms of the general reduced matrix elements $S^{j_\nu}_{L_1 p_1, \, L_2 p_2}(\omega_2)$, defined in Eq.~(\ref{S_one_electron}). By restricting ourselves to contributions of order $(\alpha Z)^2$ or lower, the angular correlation is given approximately by

\begin{eqnarray}
W^{y}(\theta) &\propto & \left|\mathcal{S}_{E1}\right|^2 \left[ \left(1  - 4 \left|\frac{\mathcal{D}_{E1}}{\mathcal{S}_{E1}} \right| \right)\left(1+\cos^2\theta \right) \right. \nonumber \\
  &-&\left. 4 \left| \frac{ \mathcal{S}_{M1}}{ \mathcal{S}_{E1}} \right| \cos\theta - 4  \left| \frac{\mathcal{S}_{E2}}{\mathcal{S}_{E1}}  \right|\cos^{3}\theta  \right]~,
\label{Acorr_ex_1}
\end{eqnarray}
where, for the sake of brevity, we  introduced the notation 
\begin{eqnarray}
\mathcal{S}_{L p} = S^{j_\nu}_{L p, \, L p}(\omega_2) + S^{j_\nu}_{L p, \, L p}(\omega_1) \nonumber \\ 
- S^{j_\nu+1}_{L p, \, L p}(\omega_2) - S^{j_\nu+1}_{L p, \, L p}(\omega_1) ~,
\end{eqnarray}
and
\begin{eqnarray}
\mathcal{D}_{L p} = 2 S^{j_\nu}_{L p, \, L p}(\omega_2) + 2 S^{j_\nu}_{L p, \, L p}(\omega_1) \nonumber \\ 
+ S^{j_\nu+1}_{L p, \, L p}(\omega_2) + S^{j_\nu+1}_{L p, \, L p}(\omega_1) ~, 
\end{eqnarray}
with $j_\nu=1/2$ for the multipoles $2E1$ and $2M1$ and  $j_\nu=3/2$ for the $2E2$  case.
 By expanding the wave-functions and energies inside the terms $S^{1/2}_{2E1}(\omega)$ and $S^{3/2}_{2E1}(\omega)$ in powers of $\alpha Z$ ($\alpha$ being the fine structure constant), we find that the term $\mathcal{D}_{E1}$ would be equal to zero, if non-relativistic wave-functions and energies were employed. Furthermore, this term scales with the atomic number as $(\alpha Z)^2$. 
 As can be observed from Eq.~\eqref{Acorr_ex_1}, the deviations of the angular correlation from the low-$Z$ regime arise from both interference between the leading multipole $2E1$ and the next higher multipoles $2M1$ and $2E2$ as well as from relativistic effects. 
 The term $\mathcal{S}_{ E1}$ is five orders of magnitude higher than the terms $\mathcal{S}_{ M1}$ and $\mathcal{S}_{ E2}$ for $Z=1$. While the term $\mathcal{S}_{ E1}$ scales as $(\alpha Z)^0$, the terms $\mathcal{S}_{ M1}$ and $\mathcal{S}_{ E2}$ scales as $ (\alpha Z)^2$, which gives rise to the asymmetry of the angular correlation for higher-$Z$ ions.  
\begin{figure*}
\centering
\includegraphics[height=8cm, width=\textwidth]{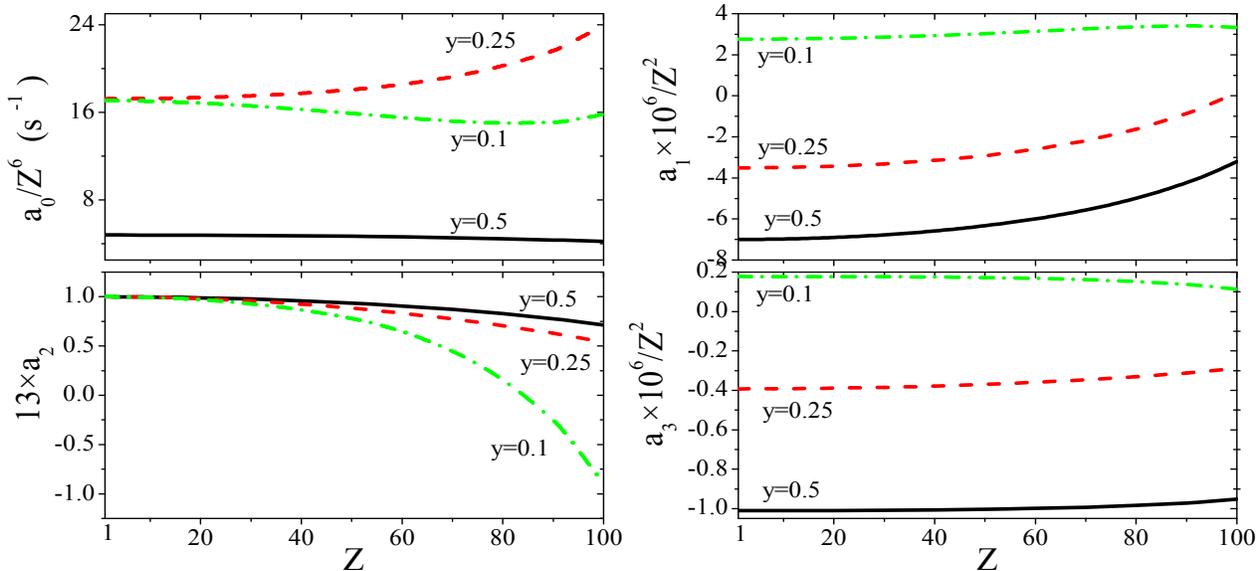} 
\caption{(Color online)  Values of the parameters $a_{i}$ defined in Eq.~(\ref{Acorr_expression}), for the $2d_{3/2}\rightarrow1s_{1/2}$ transition as function of $Z$. The different energy shares $y$ = 0.1, 0.25, 0.5 are represented by a black-solid, red-dashed and green-dot-dashed curve, respectively. With exception of $a_0$ (given in s$^{-1}$), all parameters are dimensionless. Values for parameters $b_{i}$ can be obtained from this figure (within its resolution) using the relations $b_0\approx-a_0/13$ and $b_1\approx13 a_3$.}
\label{fig3}
\end{figure*}
The expression \eqref{Acorr_ex_1} evaluated for the case of two emitted co-linear photons ($\theta=0$) has the same proportionality to the multipoles as a similar expression (18) obtained  for the case of the absorption of two co-linear photons \cite{SIS11}. 
Parameters $a_2$ and $a_4$ deviates from one and zero, respectively, for high-$Z$ ions due to terms $S^{j}_{2M1}(\omega)*S^{j'}_{2M1}(\omega')$,  $S^{j}_{2E2}(\omega)*S^{j'}_{2E2}(\omega')$ or cross multipoles not listed in Eq.~\eqref{Acorr_ex_1}, which scales as $ (\alpha Z)^4$.  These terms, along with similar ones in $a_0$, give the contribution of the multipoles $M1M1$, $E2E2$, $E1M2$ and $M2E1$ to the total decay rate. Parameters $a_{i}$ with $i>4$ depend on even higher multipoles and can be neglected even for $Z=100$.

The asymmetry of the angular correlation function was theoretically first studied by Au~\cite{Au76}  using a non-relativistic approach with the inclusion of higher (non-dipole) order multipoles. For comparison purposes, in Fig.~\ref{fig2} we also show the parameters obtained by Au for the $2s_{1/2}\rightarrow1s_{1/2}$ transition and $y=0.1,~0.25,~0.5$. For low-$Z$ ions, we found good agreement between our results and Au results in all parameters \cite{footnote}. 
On the other hand, the agreement between the two data sets becomes worst for increasing values of $Z$. As can be observed from Eq.~\eqref{Acorr_ex_1}, the first correction to the term  $a_0$ comes from relativistic effects, i. e., from the term $\mathcal{D}_{E1} $. In our work the full extent of these effects was taken into account using the Dirac theory, while in Ref.~\cite{Au76} non-relativistic wave-functions were used with a relativistic correction to the $2E1$ multipole. 

To further assess our parameters $a_0$ and $a_2$, we evaluated the total decay rate by integrating our angular correlation function in $dy$ and $d\cos\theta$ and compared it with Refs.~\cite{GoD81,SaP98}. By integrating our parametrization (\ref{Acorr_expression}) in $d\cos\theta$, only the even parameters contribute to the total decay rate. Thus, the main contribution comes from $a_0$ and $a_2$. The bottom right panel of Fig.~\ref{fig2} shows a plot of the total decay rate for several values of $Z$ and values from other references. The difference between the values of Ref.~\cite{GoD81} and Ref.~\cite{SaP98} is so small that both references can be represented by a single line. The plot in Fig.~\ref{fig2}, highlights that our results are in good agreement with the above two references. However, this is not the case for the Au values of the decay rate, which have a maximum discrepancy of 12\% for $Z=100$.

We observed good agreement between our results of the parameters $a_1$ and the respective results of Ref.~\cite{Au76} for $y=0.5$, and this would seem to indicate a good agreement of the $2M1$ multipole contribution (see Eq.~(\ref{Acorr_ex_1})). 
With decreasing energy sharing values there is less agreement, being the maximum difference at $y=0.1$ and $Z=100$ of 60\%. On the other hand, for the parameters $a_3$, there is some disagreement in the $2E2$ multipole with high-$Z$ ions at all energy shares, being the maximum difference of 35\% for $y=0.25$ and $Z=100$. There were similar discrepancy values for the $a_2$ parameter  and the maximum difference was 20\% for $Z=100$ and $y=0.1$. In the parametrization provided by Ref.~\cite{Au76}, the $a_2$ is equal to one for all energy shares and atomic numbers.
\begin{figure*}
\centering
\includegraphics[height=8cm, width=\textwidth]{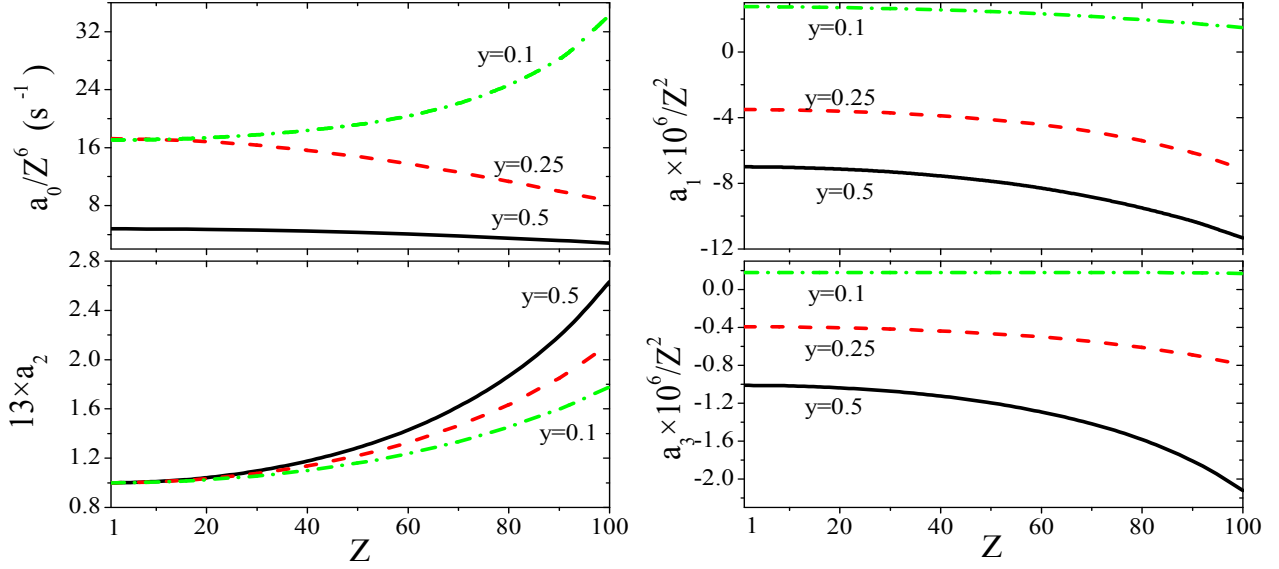}
\caption{(Color online)  Values of the parameters $a_{i}$ defined in Eq.~(\ref{Acorr_expression}), for the $2d_{5/2}\rightarrow1s_{1/2}$ transition as a function of $Z$. The different energy shares $y$ = 0.1, 0.25, 0.5 are represented by a black-solid, red-dashed and green-dot-dashed curve, respectively. With exception of $a_0$ (given in s$^{-1}$), all parameters are dimensionless.}
\label{fig4}
\end{figure*}

The asymmetry of the correlation function due to the relativistic and non-dipole contribution was also investigated by Surzhykov $et~al.$ \cite{SuK05}, who carried out a relativistic evaluation of all relevant multipoles.
Table \ref{Tab_intens} shows values for the intensity ratio between the angular correlation under 0$^{\circ}$ and 180$^{\circ}$ obtained in the present work, as well as by Au~\cite{Au76} and Surzhykov $et~al.$ \cite{SuK05}.
As can be observed, good agreement is found for the values presented by Surzhykov $et~al.$ \cite{SuK05}.

%
\begin{table}
\caption{\label{Tab_intens} Intensity ratio between the angular correlation between 0$^{\circ}$ and 180$^{\circ}$ for several values of the atomic number and equal energy sharing ($y=0.5$).  Comparison between
  the values obtained in this work and other theoretical
  values. 
}
\begin{ruledtabular}
\begin{tabular}{llll}
& \multicolumn{3}{c}{$W^{0.5}(180^{\circ})/ W^{0.5}(0^{\circ})$} \\ 

\hline
  & $Z=1$& $Z=54$ & $Z=92$   \\ 
\cline{2-4}
Surzhykov $et~al.$ \cite{SuK05}& 1.000 & 1.038          & 1.124           \\
Au \cite{Au76} &   1.000  & 1.038&     1.131     			\\
 This work &  1.000  & 1.038       &      1.123    	\\

\end{tabular}
\end{ruledtabular}

\end{table}
Overall, we noticed that deviations from the formula $W^y(\theta)=a_0(1 + \cos^2\theta)$ start playing the role of some percent from hydrogen-like Tin ion onwards ($Z\gtrsim 50$).

As with the angular correlation function, the non-dipole and relativistic effects on the degree of linear polarization function can also be estimated by expressing this function in terms of the reduced matrix elements. The result of this procedure is given by
\begin{widetext}
\begin{equation}
P_L(\theta) \approx \left| \frac{ -\left(1  - 4 \left|\frac{\mathcal{D}_{E1}}{\mathcal{S}_{E1}}\right|\right)\sin^2\theta \left(1- 4 \left|\frac{\mathcal{S}_{E2}}{\mathcal{S}_{E1}}\right|\cos \theta \right)}{   \left(1  - 4 \left|\frac{\mathcal{D}_{E1}}{\mathcal{S}_{E1}}\right|\right)\left(1+\cos^2\theta\right)
 - 4\left| \frac{ \mathcal{S}_{M1}}{ \mathcal{S}_{E1}}\right| \cos\theta   
  - 4 \left|\frac{\mathcal{S}_{E2}}{\mathcal{S}_{E1}}\right|\cos^{3}\theta }  \right|~.
\label{PL_ex_1}
\end{equation}
\end{widetext}

Likewise, the numerator of the degree of linear polarization (\ref{Pl_par}) in $2s_{1/2}\to1s_{1/2}$ transitions can be well described by $a_0\sin^2 \theta $, so that $P_L(\theta)= \sin^2 \theta/(1+\cos^2 \theta)$ \cite{SRI09}, as long as the atomic number is relatively small. 

Eq.~\eqref{PL_ex_1} highlights that $a_0= -b_0$ and $a_3=b_1$ if multipoles lower than three were employed and also in this case $b_0$ and $b_1$ could be obtained from Fig.~\ref{fig2}. 
The maximum degree of polarization corresponds to $\theta=90^{\circ}$ ($P_L(90^{\circ})=|b_0/a_0|$). This quantity is almost independent of non-dipole and relativistic effects with explicit values of $1$ for $y=0.5$ and $0.9$ for $y=0.1$. 
As in the case of $a_{4}$, parameters  $b_{4}$ depends on multipole contributions of order of $(\alpha Z)^4$, which are not shown in Eq.~\eqref{PL_ex_1}. We found that $a_4 \approx -b_4$ for all the isoelectronic sequence. Moreover, like the case of $a_{i}$, parameters  $b_{i}$ with $i>4$ depend on higher multipoles and therefore can be neglected.

As shown in Fig.~\ref{fig2}, the deviations from both the angular correlation and the degree of linear polarization present a non-trivial dependence on the energy sharing that characterizes the decay: 
$a_{i=0,2 , 4}$ are greater in magnitude for higher energy shares, while $a_{i=1,3}$ and $b_{i=0,1,4}$ are smaller with increasing values of energy shares. 

Other transitions in which a two-photon emission is observed are the $3d_{3/2}\to1s_{1/2}$ and $3d_{5/2}\to1s_{1/2}$ \cite{MoD04, MoD042}.
Figs.~\ref{fig3} and \ref{fig4} report the coefficients $a_i$ and $b_i$ as obtained for the two-photon transitions $3d_{3/2}\to1s_{1/2}$ and $3d_{5/2}\to1s_{1/2}$, respectively.
Similar to Fig.~\ref{fig1},  deviations from the non-relativistic formula $W^y(\theta)\approx a_0(1 + \cos^2\theta/13)$ \cite{Flore} are of the order of some percent from $Z\approx 50$ onwards. Also in the lower $Z$ regime the degree of linear polarization is given by $P_L(\theta)=\sin^2 \theta/(13+\cos^2 \theta)$. 

As for the $a_i$ case, parameters $b_i$ of the transition $3d_{3/2}\to1s_{1/2}$ can be obtained from Fig.~\ref{fig3}  using the following relations $b_0\approx-a_0/13$ and  $b_1\approx 13 a_3$. Like the case of the transition $2s_{1/2}\to1s_{1/2}$, relativistic and non-dipole effects do not give a sizable contribution to the maximum degree of polarization ($P_L(90^{\circ})=1/13$) of the first photon emitted by the transition $3d_{3/2}\to1s_{1/2}$. 
However, this is not case of the transition $3d_{5/2}\to1s_{1/2}$, where $P_L(90^{\circ})$ increases by 20\% from neutral hydrogen to H-like uranium for $y=0.5$.
%

Tables \ref{tab_1} -- \ref{tab_3} in the Appendix \ref{appen_tab} provide the values of the parameters $a_i$ and $b_i$ with $i<5$ for the transitions considered in this work for a few atomic numbers and for the case of equal energy sharing. The values of parameters not shown in the Figs.~\ref{fig3} and \ref{fig4}, can be found in these tables.

The angular correlation and the degree of linear polarization expressed as Eqs. (\ref{Acorr_expression}) and (\ref{Pl_par}), together with the parameters in Figs. ~\ref{fig2}-~\ref{fig4} provide an accurate fit model for any future experiments that involves two-photon polarization, for example, the parity non-conservation mixing coefficient measurement \cite{PNC}.

%
\section{Summary}
\label{sum}

We analyzed the angular correlation and the degree of linear polarization for the radiation emitted in two-photon decay of hydrogen-like ions for the transitions $2s_{1/2}\rightarrow1s_{1/2}$, $3d_{3/2}\rightarrow1s_{1/2}$ and $3d_{5/2}\rightarrow1s_{1/2}$. These two physical quantities were expanded in $\cos\theta$-polynomials and the coefficients were plotted for the entire isoelectronic sequence ($1\le Z\le100$). 
By restricting ourselves to the first two multipoles of the photon field expansions,  these coefficients were written in terms of the reduced amplitude of the process. Overall, we have shown that the coefficients that deviate from non-relativistic formulae, begin to be some percent from  approximately hydrogen-like Tin ion onwards ($Z\gtrsim 50$).

When available, a comparison with references was performed in order to verify these present results.

The parametrization of the angular correlation and the degree of linear polarization presented in this work could be exploited in future experiments aim at measurering the angular and polarization properties of the radiation emitted in two-photon decay of atoms or ions.

%
\begin{acknowledgments}

This research was supported in part by FCT -- Funda{\c c}\~ao para a Ci\^encia e a Tecnologia (Portugal), through the projects No. PEstOE/FIS/UI0303/2011 and PTDC/FIS/117606/2010, financed by the European Community Fund FEDER through the COMPETE Ð Competitiveness Factors Operational Programme, by the French-Portuguese collaboration (PESSOA Program, contract No. 441.00), by the Ac\c{c}\~oes Integradas Luso-Francesas (contract No. F-11/09) and by the Programme Hubert Curien and Ac\c{c}\~oes Integradas Luso-Alem\~as (contract No. 20022VB and A-19/09).
This research was partly supported by the Helmholtz Alliance HA216/EMMI. Laboratoire Kastler Brossel is \supercomas{Unit\'e Mixte de Recherche n$^\circ$ 8552} of \'Ecole Normale Sup\'erieure, CNRS and Universit\'e Pierre et Marie Curie.
P.A. acknowledges the support of the FCT, under contract No. SFRH/BD/37404/2007 and German Research Foundation (DFG) within the Emmy Noether program under contract No. TA 740 1-1.
A. S. and F. F. acknowledge support from the Helmholtz Gemeinschaft and GSI under project No. VH-NG-421 and from the Deutscher Akademischer Austauschdienst (DAAD) under project No. 0813006. S. F. acknowledges the support by the FiDiPro program of the Finish Academy.
 F. F. acknowledges the support by the Research Council for Natural Sciences and Engineering of the Academy of Finland.

\end{acknowledgments}


\appendix
\section{Tables of the parameters}
\label{appen_tab}

The values for the parameters $a_{i}$ and $b_{i}$ 
are shown, for a few atomic numbers and equal energy sharing. $b_{2}$ and $b_{3}$ are not shown since the relations $b_2=-(1+b_4)$  and $b_3=-b_1$ hold true.\\
The considered transitions are
$2s_{1/2}\rightarrow1s_{1/2}$, $3d_{3/2}\rightarrow1s_{1/2}$ and $3d_{5/2}\rightarrow1s_{1/2}$, which are displayed in Tables \ref{tab_1}, \ref{tab_2} and \ref{tab_3}, respectively. 

\onecolumngrid

\begin{table}[h!]
\caption{Values of the parameters $a_i$ and $b_i$ defined in Eqs.~(\ref{Acorr_expression}) and (\ref{Pl_par}), for the transition
$2s_{1/2} \rightarrow 1s_{1/2}$. The energy sharing is fixed at $y=0.5$. 
All parameters are dimensionless, with the exception of $a_0$ and $b_0$, which are given in s$^{-1}$.
}
\begin{ruledtabular}
\begin{tabular}{p{1cm}p{2cm}p{2cm}p{2cm}p{2cm}p{2cm}p{2cm}p{2cm}p{2cm}}
$Z$    &     $a_{0}/Z^6 $    &    $a_{1} \times 10^6 / Z^2$  &    $a_{2} $   &    $a_{3} \times 10^6 / Z^2$  &    $a_{4}\times 10^{12} / Z^4$ & $b_{0}/Z^6$  &   $b_{1} \times 10^6 / Z^2$     &    $b_{4}\times 10^{12} / Z^4$  \\
\hline
1	&	3.9942	&	-6.7348	&	1.0000	&	-5.4827	&	7.5149	&	-3.9942	&	-5.4827	&	-7.5149	\\
4	&	3.9935	&	-6.7378	&	1.0000	&	-5.4826	&	7.5147	&	-3.9935	&	-5.4826	&	-7.5147	\\
8	&	3.9911	&	-6.7472	&	1.0000	&	-5.4824	&	7.5142	&	-3.9911	&	-5.4824	&	-7.5142	\\
12	&	3.9871	&	-6.7630	&	1.0000	&	-5.4820	&	7.5132	&	-3.9871	&	-5.4821	&	-7.5132	\\
16	&	3.9816	&	-6.7851	&	1.0000	&	-5.4815	&	7.5119	&	-3.9816	&	-5.4816	&	-7.5119	\\
20	&	3.9744	&	-6.8135	&	1.0000	&	-5.4808	&	7.5101	&	-3.9744	&	-5.4809	&	-7.5102	\\
30	&	3.9494	&	-6.9118	&	0.9998	&	-5.4781	&	7.5037	&	-3.9491	&	-5.4788	&	-7.5040	\\
40	&	3.9142	&	-7.0488	&	0.9993	&	-5.4736	&	7.4941	&	-3.9132	&	-5.4757	&	-7.4951	\\
50	&	3.8683	&	-7.2232	&	0.9984	&	-5.4664	&	7.4807	&	-3.8658	&	-5.4716	&	-7.4830	\\
54	&	3.8468	&	-7.3029	&	0.9978	&	-5.4625	&	7.4740	&	-3.8434	&	-5.4697	&	-7.4771	\\
60	&	3.8111	&	-7.4323	&	0.9966	&	-5.4553	&	7.4625	&	-3.8058	&	-5.4665	&	-7.4672	\\
70	&	3.7419	&	-7.6704	&	0.9935	&	-5.4385	&	7.4381	&	-3.7319	&	-5.4602	&	-7.4466	\\
80	&	3.6591	&	-7.9268	&	0.9885	&	-5.4136	&	7.4063	&	-3.6418	&	-5.4528	&	-7.4202	\\
90	&	3.5614	&	-8.1788	&	0.9808	&	-5.3766	&	7.3640	&	-3.5332	&	-5.4436	&	-7.3849	\\
92	&	3.5402	&	-8.2248	&	0.9788	&	-5.3671	&	7.3535	&	-3.5093	&	-5.4414	&	-7.3759	\\
100	&	3.4474	&	-8.3798	&	0.9692	&	-5.3215	&	7.3078	&	-3.4036	&	-5.4321	&	-7.3363	\\
\end{tabular}
\end{ruledtabular}
\label{tab_1}
\end{table}

\begin{table}[h!]
\caption{Same as Tab.\ti\ref{tab_1} but for
 $3d_{3/2} \rightarrow 1s_{1/2}$ transition.}
\begin{ruledtabular}
\begin{tabular}{p{1cm}p{1.5cm}p{2cm}p{1.5cm}p{2cm}p{2cm}p{2cm}p{2.5cm}p{2.5cm}}
$Z$    &     $a_{0}/Z^6 $    &   $a_{1} \times 10^6 / Z^2$    &    $13 \times a_{2} $   &    $a_{3} \times 10^6 / Z^2$   &    $a_{4}\times 10^{12} / Z^4$
           &     $13\times b_{0}/Z^6$         &     $b_{1} \times 10^6 / (13 Z^2)$      &    $b_{4}\times 10^{12} / (13 Z^4)$\\
\hline
1	&	4.7852	&	-6.9947	&	1.0000	&	-1.0110	&	3.3220	&	-4.7852	&	-1.0110	&	-3.3220	\\
4	&	4.7846	&	-6.9911	&	0.9996	&	-1.0110	&	3.3228	&	-4.7839	&	-1.0112	&	-3.3233	\\
8	&	4.7824	&	-6.9795	&	0.9984	&	-1.0108	&	3.3254	&	-4.7797	&	-1.0118	&	-3.3273	\\
12	&	4.7788	&	-6.9600	&	0.9963	&	-1.0106	&	3.3297	&	-4.7726	&	-1.0128	&	-3.3340	\\
16	&	4.7738	&	-6.9326	&	0.9934	&	-1.0103	&	3.3358	&	-4.7628	&	-1.0143	&	-3.3435	\\
20	&	4.7673	&	-6.8969	&	0.9897	&	-1.0099	&	3.3438	&	-4.7501	&	-1.0162	&	-3.3559	\\
30	&	4.7443	&	-6.7704	&	0.9768	&	-1.0084	&	3.3721	&	-4.7058	&	-1.0228	&	-3.3997	\\
40	&	4.7111	&	-6.5852	&	0.9587	&	-1.0062	&	3.4139	&	-4.6431	&	-1.0324	&	-3.4638	\\
50	&	4.6666	&	-6.3323	&	0.9352	&	-1.0031	&	3.4713	&	-4.5612	&	-1.0454	&	-3.5513	\\
54	&	4.6453	&	-6.2092	&	0.9243	&	-1.0016	&	3.4994	&	-4.5229	&	-1.0516	&	-3.5939	\\
60	&	4.6090	&	-5.9978	&	0.9061	&	-0.9988	&	3.5479	&	-4.4590	&	-1.0623	&	-3.6671	\\
70	&	4.5365	&	-5.5608	&	0.8708	&	-0.9928	&	3.6490	&	-4.3353	&	-1.0838	&	-3.8179	\\
80	&	4.4453	&	-4.9896	&	0.8282	&	-0.9844	&	3.7829	&	-4.1876	&	-1.1113	&	-4.0148	\\
90	&	4.3327	&	-4.2339	&	0.7762	&	-0.9721	&	3.9615	&	-4.0146	&	-1.1461	&	-4.2737	\\
92	&	4.3080	&	-4.0543	&	0.7645	&	-0.9688	&	4.0037	&	-3.9775	&	-1.1541	&	-4.3343	\\
100	&	4.1957	&	-3.2113	&	0.7113	&	-0.9530	&	4.2046	&	-3.8159	&	-1.1905	&	-4.6199	\\

\end{tabular}
\end{ruledtabular}
\label{tab_2}
\end{table}

\begin{table}[h!]
\caption{Same as Tab.\ti\ref{tab_1} but for
$3d_{5/2} \rightarrow 1s_{1/2}$ transition.}
\begin{ruledtabular}
\begin{tabular}{p{1cm}p{1.5cm}p{2cm}p{1.5cm}p{2cm}p{2cm}p{2cm}p{2.5cm}p{2.5cm}}

$Z$    &     $a_{0}/Z^6 $    &    $a_{1} \times 10^6 / Z^2$   &    $13 \times a_{2}$   &    $a_{3} \times 10^6 / Z^2$   &   $a_{4}\times 10^{12} / Z^4$
&     $13\times b_{0}/Z^6$         &     $b_{1} \times 10^6 / (13 Z^2)$      &    $b_{4}\times 10^{12} / (13 Z^4)$\\
\hline
1	&	4.7851	&	-6.9953	&	1.0001	&	-1.0111	&	3.3221	&	-4.7852	&	-1.0110	&	-3.3219	\\
4	&	4.7820	&	-7.0002	&	1.0016	&	-1.0121	&	3.3243	&	-4.7848	&	-1.0111	&	-3.3224	\\
8	&	4.7722	&	-7.0158	&	1.0066	&	-1.0153	&	3.3317	&	-4.7832	&	-1.0113	&	-3.3240	\\
12	&	4.7558	&	-7.0420	&	1.0149	&	-1.0207	&	3.3439	&	-4.7805	&	-1.0117	&	-3.3266	\\
16	&	4.7329	&	-7.0788	&	1.0266	&	-1.0283	&	3.3612	&	-4.7766	&	-1.0123	&	-3.3305	\\
20	&	4.7036	&	-7.1265	&	1.0418	&	-1.0381	&	3.3837	&	-4.7712	&	-1.0132	&	-3.3356	\\
30	&	4.6018	&	-7.2946	&	1.0960	&	-1.0733	&	3.4639	&	-4.7503	&	-1.0165	&	-3.3553	\\
40	&	4.4601	&	-7.5374	&	1.1756	&	-1.1251	&	3.5822	&	-4.7148	&	-1.0223	&	-3.3877	\\
50	&	4.2790	&	-7.8629	&	1.2846	&	-1.1964	&	3.7458	&	-4.6583	&	-1.0317	&	-3.4382	\\
54	&	4.1957	&	-8.0187	&	1.3378	&	-1.2313	&	3.8261	&	-4.6282	&	-1.0369	&	-3.4651	\\
60	&	4.0593	&	-8.2827	&	1.4294	&	-1.2916	&	3.9653	&	-4.5729	&	-1.0465	&	-3.5145	\\
70	&	3.8024	&	-8.8127	&	1.6188	&	-1.4175	&	4.2570	&	-4.4493	&	-1.0689	&	-3.6274	\\
80	&	3.5091	&	-9.4751	&	1.8665	&	-1.5844	&	4.6467	&	-4.2759	&	-1.1022	&	-3.7938	\\
90	&	3.1821	&	-10.2988	&	2.1932	&	-1.8090	&	5.1749	&	-4.0418	&	-1.1513	&	-4.0390	\\
92	&	3.1134	&	-10.4854	&	2.2706	&	-1.8629	&	5.3018	&	-3.9874	&	-1.1635	&	-4.1003	\\
100	&	2.8250	&	-11.3217	&	2.6331	&	-2.1191	&	5.9100	&	-3.7368	&	-1.2236	&	-4.4060	\\

\end{tabular}
\end{ruledtabular}
\label{tab_3}
\end{table}

\pagebreak

\twocolumngrid

\pagebreak
\twocolumngrid


\newpage
\begin{thebibliography}{99}
\bibitem{GoM31} M.~Goeppert-Mayer, Ann. Phys. (Leipzig) {\bf 9}, 273 (1931).

\bibitem{BrT40} G.~Breit and E.~Teller, Astrophys.J. {\bf 91}, 215 (1940).

\bibitem{Au76}  C.~K.~Au, Phys. Rev. A {\bf 14}, 531 (1976).

\bibitem{GoD81} S. P. Goldman and G.W.F. Drake, Phys. Rev. A {\bf 24}, 183 (1981).

\bibitem{DuB93} R.~W.~Dunford, H.~G.~Berry, S.~Cheng, E.~P.~Kanter, C.~Kurtz, B.~J.~Zabransky, A.~E.~Livingston, L.~J.~Curtis,
Phys. Rev. A {\bf 48}, 1929 (1993).

\bibitem{AlA97} R.~Ali, I.~Ahmad, R.~W.~Dunford, D.~S.~Gemmell, M.~Jung, E.~P.~Kanter, P.~H.~Mokler, H.~G.~Berry,
A.~E.~Livingston, S.~Cheng, and L.~J.~Curtis, Phys. Rev. A {\bf 55}, 994 (1997).

\bibitem{ScM99} H.~W.~Sch\"affer, P.~H.~Mokler, R.~W.~Dunford, C.~Kozhuharov, A. Kr\"amer, A.~E.~Livingston,
T.~Ludziejewski, H.-T.~Prinz, P.~Rymuza, L.~Sarkadi, Z.~Stachura, Th.~St\"ohlker, P.~Swiat, and A.~Warczak,
Phys. Lett. A {\bf 260}, 489 (1999).

\bibitem{MoD04} P.~H.~Mokler and R.~W.~Dunford, Phys. Scr. {\bf 69}, C1 (2004).

\bibitem{MoD042}  P. H. Mokler, H. W. Sch\"affer, and R. W. Dunford, Phys. Rev. A
{\bf 70}, 032504 (2004).

\bibitem{IlU06} K.~Ilakovac, M.~Uroi\`c, M.~Majer, S.~Pasi\`c, and B.~Vukovi\'c, Rad. Phys. Chem. {\bf 75}, 1415 (2006).

                

\bibitem{ToL90} X.~M.~Tong, J.~M.~Li, L.~Kissel, and R.~H.~Pratt, Phys. Rev. A {\bf 42}, 1442 (1990).

\bibitem{DeJ97} A.~Derevianko and W.~R.~Johnson, Phys. Rev. A {\bf 56}, 1288 (1997).

\bibitem{SaP98} J. P. Santos, F. Parente, and P. Indelicato, Eur. Phys. J. D {\bf 3}, 43 (1998).

\bibitem{KuT09} A.~Kumar, S.~Trotsenko, A.~V.~Volotka, D.~Bana\'s, H.~F.~Beyer, H.~Br\"auning, A.~Gumberidze,
S.~Hagmann, S.~Hess, C.~Kozhuharov, R.~Reuschl, U.~Spillmann, M.~Trassinelli, G.~Weber, and Th.~St\"ohlker,
Eur. Phys. J. Special Topics {\bf 169}, 19 (2009).

\bibitem{TrS09}S. Trotsenko,  A. Kumar,  A. V. Volotka,  D. Bana\'{s},  H. F. Beyer,  H. Br\"auning,  S. Fritzsche,  A. Gumberidze,  S. Hagmann, 
S. Hess,  P. Jagodzi\'{n}ski,  C. Kozhuharov,  R. Reuschl,  S. Salem,  A. Simon,  U. Spillmann,  M. Trassinelli,  L. C. Tribedi,  G. Weber,  D. Winters, 
Th. St\"ohlker, Phys. Rev. Lett. {\bf 104}, 033001 (2010).

\bibitem{pdb1985}  W. Perrie, A. J. Duncan, H. J. Beyer, H. Kleinpoppen, Phys. Rev. Lett. {\bf 74}, 1790 (1985). 

\bibitem{ssi1989} A. Sch\"{a}fer, G. Soff, P. Indelicato, B. M\"{u}ller, W. Greiner, Phys. Rev. A {\bf 40}, 7362 (1989).

\bibitem{KlD97} H.~Kleinpoppen, A.~J.~Duncan, H.--J.~Beyer, and Z.~A.~Sheikh, Phys. Scr. {\bf T72}, 7 (1997).
                
\bibitem{FrI05} S.~Fritzsche, P.~Indelicato, and Th.~St\"ohlker, J. Phys. B: At. Mol. Opt. Phys. {\bf 38}, S707 (2005).

\bibitem{PNC}   F.~Fratini, S.~Trotsenko, S.~Tashenov, Th.~St\"ohlker and A.~Surzhykov, Phys. Rev. A {\bf 83}, 052505 (2011). 

\bibitem{QuantumInf} F. Fratini, M. C. Tichy, Th. Jahrsetz, A. Buchleitner, S. Fritzsche, 
and A. Surzhykov, Phys. Rev. A {\bf83}, 032506 (2011).

\bibitem{SIS11} A. Surzhykov, P. Indelicato, J. P. Santos, P. Amaro, S. Fritzsche, Phys. Rev. A {\bf 84}, 022511 (2011).

\bibitem{RSI67} P.~S.~Shaw, U.~Arp, A.~Henins and S.~Southworth, Rev. Sci. Instrum. {\bf 67}, 3362 (1996).

\bibitem{TaS06} S.~Tashenov, Th.~St\"ohlker, D.~Bana\'s, K.~Beckert, P.~Beller, H.~F.~Beyer, F.~Bosch, S.~Fritzsche, A.~Gumberidze,
S.~Hagmann, C.~Kozhuharov, T.~Krings, D.~Liesen, F.~Nolden, D.~Protic, D.~Sierpowski, U.~Spillmann, M.~Steck, A.~Surzhykov,
Phys. Rev. Lett. {\bf 97}, 223202 (2006).

\bibitem{StS07} Th.~St\"ohlker, U.~Spillmann, D.~Bana\'s, H.~F.~Beyer, J.~Cl.~Dousse, S.~Chatterjee, S.~Hess, C.~Kozhuharov,
M.~Kavcic, T.~Krings, D.~Proti\'c, R.~Reuschl, J.~Szlachetko, S.~Tashenov, and S.~Trotsenko, J. Phys.: Conf. Ser. {\bf 58}, 411 (2007).



\bibitem{SuK05} A. Surzhykov, P. Koval, and S. Fritzsche, Phys. Rev. A {\bf 71}, 022509 (2005).  


\bibitem{SuV10} A.~Surzhykov, A.~Volotka, F.~Fratini, J.~P.~Santos, P.~Indelicato, G.~Plunien, Th.~St\"ohlker and S.~Fritzsche,
Phys. Rev. A {\bf 81}, 042510 (2010).

\bibitem{Hyperfine} F. Fratini and A. Surzhykov, Hyp. Int. {\bf 199}, 85 (2011).


\bibitem{Rose} M.~Rose, Elementary Theory of Angular Momentum (Wiley, New York, 1957).

\bibitem{PRL29} W.~R.~Johnson, Phys. Rev. Lett. {\bf 29}, 1123 (1972).

\bibitem{JPB31} A.~Maquet, V.~V\'eniard and T.~A.~Marian, J. Phys. B. {\bf 31}, 3743 (1998).


\bibitem{JPA24}	R.~A.~Swainson and G.~W.~F.~Drake, J. Phys. A: Math. Gen. {\bf 24}, 95 (1991);\\
					R.~A.~Swainson and G.~W.~F.~Drake, J. Phys. A: Math. Gen. {\bf 24}, 79 (1991);\\
					R.~A.~Swainson and G.~W.~F.~Drake, J. Phys. A: Math. Gen. {\bf 24}, 1801 (1991).



\bibitem{SaJ96} J.~Sapirstein and W.~R.~Johnson, J. Phys. B {\bf 29}, 5213 (1996).

\bibitem{ShT04} V.~M.~Shabaev, I.~I.~Tupitsyn, V.~A.~Yerokhin, G.~Plunien,
and G.~Soff, Phys. Rev. Lett. {\bf 93}, 130405 (2004).      

\bibitem{Balashov} V.~V.~Balashov, A.~N.~Grum-Grzhimailo, N. M. Kabachnik, 
{\it Polarization and Correlation Phenomena in Atomic Collisions} (Springer-Verlag , 2000).


\bibitem{asp2009} P. ~Amaro, J.~P.~ Santos,  F.~Parente, A.~Surzhykov, P.~Indelicato,  Phys. Rev. A {\bf79}, 062504 (2009).

 \bibitem{sis2011}A.~Surzhykov,  P.~Indelicato,  J.~P.~ Santos,  P. ~Amaro,  S.~Fritzsche, Phys. Rev. A {\bf84}, 022511 (2011).
 
 \bibitem{saf2012}L.~Safari, P.~Amaro, S.~Fritzsche, J.~P.~ Santos, F.~ Fratini, Phys. Rev. A {\bf85}, 043406 (2012).
 
\bibitem{asp2011} P. ~Amaro, A. ~Surzhykov, F. ~Parente, P. ~Indelicato, and J. ~P.
Santos, J. Phys. A {\bf44}, 245302 (2011).

\bibitem{Klarsfeld}	S. Klarsfeld, Phys. Lett. {\bf30A} 382 (1969).

\bibitem{yang} C. N. Yang, J. Phys. A {\bf74}, 764 (1948).

\bibitem{footnote} We noticed that  the agreement shown for the parameters $a_1$ and $a_3$ was only possible considering Eq.~(16) of Ref.~\cite{Au76} not having misplaced \supercomas{2} in the second term.

\bibitem{SRI09} A. Surzhykov, T. Radtke, P. Indelicato, S. Fritzsche, Eur. Phys. J. Special Topics {\bf 169}, 29 (2009).

\bibitem{Flore}V. Florescu, Phys. Rev. A {\bf30}, 2441 (1984)



               


\end{thebibliography}
\end{document}